\documentclass[aps,prb,twocolumn,showpacs]{revtex4}
\usepackage{epsf}
\newcommand{\A}{{\cal{A}}}
\newcommand{\B}{{\cal{B}}}

\begin{document}

       \title{Remnant superfluid collective phase oscillations in\\ 
              the normal state of systems with resonant pairing}
      \author{T.\ Domanski$^{1}$, J.\ Ranninger$^{2}$}
\affiliation{$^{1}$  Institute of Physics, M.\ Curie Sk\l odowska 
             University, 20-031 Lublin, Poland\\
	     $^{2}$Centre de Recherches sur les Tr\`es Basses 
	     Temp\'eratures CNRS associ\'e \`a l'Universit\'e 
	     Joseph Fourier, BP 166, 38-042 Grenoble C\'edex 9, France}

\begin{abstract}
The signature of superfluidity in bosonic systems is a sound wave-like
spectrum of the single particle excitations which in the case of strong
interactions is roughly temperature independent. In fermionic systems,
where fermion pairing  arises as a resonance phenomenon between free
fermions and paired fermionic states (examples are: the atomic gases
of $^{6}$Li or $^{40}$K controlled by a Feshbach resonance, polaronic
systems in the intermediary coupling regime, $d$-wave hole pairing in
the strongly correlated Hubbard system), remnants of such superfluid
characteristics are expected to be visible in the normal state.
The single particle excitations maintain there a sound wave like
structure for wave vectors above a certain  $q_{min}(T)$ where they
practically coincide there with the spectrum of the superfluid phase for
$T<T_{c}$. Upon approaching the transition from above this region
in $q$-space extends down to small momenta, except for a narrow
region around $q=0$ where such modes change into damped free particle
like excitations.
\end{abstract}
\maketitle

\section{Introduction}
Approaching the transition to a superconducting or superfluid state from 
above, one can (under certain conditions) observe incipient macroscopic 
features which are caused by the emergence of an  order parameter. In 
classical superconductors such features, related to spatial order parameter 
fluctuations, are restricted to only an extremely narrow temperature 
region around the superconducting critical temperature $T_c$, and in practice 
are hard to detect at all. Such fluctuations however are visible in systems 
with real space pairing or, more generally, when the overlap between the 
pair wave functions is small and we are in the crossover regime between 
a BCS type superfluidity of Cooperons and a superfluid phase of tightly 
bound Fermions which behave as bosons. Remnants of superfluidity, 
sometimes termed {\it localized superfluidity}, above $T_{c}$ have been 
observed \cite{Glyde-04} in form of finite range phase correlations in  
purely bosonic systems such as liquid $^{4}$He in porous media of vicors 
and aerogels, with a characteristic disorder and confinement. Similar 
features have been seen for  Fermionic systems such as $^{3}$He in 
aerogels \cite{Bunkov-00} and superconducting hetero structures 
\cite{Christiansen-02}. Solution to the theoretical questions raised 
in this connection lies in a formulation capable of describing on equal 
footing a BCS type superconductivity in a system of weakly coupled fermions  
and a Bose Einstein condensation (BEC) of strongly bound fermion pairs. 
Early attempts to do that go go back to the work of Leggett \cite{Leggett-80} 
and Nozi\`eres and Schmitt-Rink \cite{NSR-85} and rely on cross-over 
scenarios  where electron pairing is given by some unspecified effective 
attraction between them.

Fermionic systems where the binding between Fermions comes about from 
an  exchange interaction between free itinerant Fermions and two-Fermion 
bound states, present a different scenario to examine the cross-over 
regime between a BCS type  superfluidity and a condensed states of tightly 
bound pairs. Such systems have moreover the advantage that sometimes, in real 
systems, the cross-over can be tuned experimentally. 
An example for such scenarios are Many Polaron systems in the intermediary 
coupling regime where free itinerant electrons  engage in  a resonant 
scattering process with weakly bound bipolaronic states when their 
respective  energy difference is small\cite{Ranninger-02}. This leads 
to long lived electron pairs which ultimately can condensate. An other 
example, now widely studied in the literature in connection with their 
condensation\cite{atomgas-SF}, are gases of fermionic atomic  
(such as $^{6}$Li and $^{40}$K atoms)  which can be brought into such 
resonant fermionic pair states via a so called Feshbach resonance 
mechanism\cite{Timmermans-99} which involves hyperfine spin-flip processes 
between the nuclear and the electronic spins of the atoms together with 
their molecular counterparts. Finally, also in the highly debated 
scenarios for the high temperature superconductors (HTSC) resonant 
pairing  between $d$-wave holes has been invoked. There, it has been 
suggested that such pairing  arises from an exchange between itinerant 
holes and bound hole pairs in plaquette RVB states on finite 
clusters\cite{Auerbach-02}. 

In all those systems resonant pairing leads to long lived electron pairs 
which ultimately are driven into a superfluid phase. Furthermore such systems 
are characterized by a strongly interdependent dynamics of single- and 
two-particle excitations which, upon approaching and passing through 
the superconducting  phase transition,  simultaneously undergo qualitative 
changes. Thus, the opening of a pseudogap in the single particle spectrum,  
when $T_c$ is approached from above, occurs concomitantly with a change-over 
from single particle fermionic transport to one ensured by bosonic molecular 
entities \cite{Devillard-00}. The observed transient Meissner effect
\cite{Corson-99} and a Nernst effect \cite{Ong-00} in the normal phase 
in HTSC can be considered to be signatures of that. In the atomic gases 
the physics is more involved because of the strong inhomogeneous character 
of those trapped gases, leading to radial and axial breathing modes 
\cite{Collmode} instead of the usual sound-wave like excitation spectrum 
known in translational invariant homogeneous superfluids. Nevertheless, 
corresponding manifestation of superfluid fluctuations in the  normal 
state should also be expected in those systems.

In this paper we shall analyze the molecular (and/or fermion-pair) excitation 
spectrum of such a general class of systems which can be described in terms of 
resonating pairs of fermions and discuss how, on a finite length scale, 
superfluid phase fluctuations can emerge upon approaching  $T_{c}$ from above. 
We restrict ourselves here to the study of homogeneous systems, leaving the 
more complex structures to be expected for remnant collective modes in 
inhomogeneous atomic gases in optical traps to a future work. The simplest, 
and generally adopted approach to study such systems, is on 
the basis of a phenomenological Boson-Fermion model.

Pairing in such a model can be viewed as the Andreev scattering processes 
between itinerant carriers and bosonic bound pairs on small clusters. One 
then is faced on one hand with local intra-cluster phase correlations between 
pairs of itinerant fermions and bosonic bound Fermion pairs and on the other
hand with non-local inter-cluster phase correlations \cite{Cuoco-03,Cuoco-04}. 
The first ones play the role of local density fluctuations and the second ones 
of effective inter-site Josephson coupling. This physics, which is an intrinsic 
ingredient of the various representative examples cited above and which are 
effectively realized in nature, is qualitatively different from that of 
the standard cross-over scenarios based on effective attractive inter-particle 
interactions. It leads to novel features such as superfluid-insulator 
transitions, and lets one envisage the possibility of normal state bose 
metals and exotic elementary as well as collective excitations which remain 
to be fully explored \cite{Cuoco-04}.

\section{The model} 
The following boson fermion model (BFM) Hamiltonian for resonant pairing
\begin{eqnarray}
H &=& \sum_{{\bf k},\sigma} \varepsilon_{\bf k} 
c_{{\bf k}\sigma}^{\dagger} c_{{\bf k}\sigma} 
+ v \sum_{{\bf k},{\bf q}} \left(  b_{\bf q}^{\dagger} 
c_{{\bf q}-{\bf k}\downarrow}c_{{\bf k}\uparrow} 
+ \mbox{h.c.} \right) \nonumber \\
&+& \sum_{\bf q} \left( E_{\bf q} + 2\nu \right) 
b_{\bf q}^{\dagger} b_{\bf q} .
\label{BF}
\end{eqnarray}
is currently employed in  studies of the above mentioned systems. 
The operators $c_{{\bf k}\sigma}^{\dagger}$ ($c_{{\bf k}\sigma}$) 
correspond, according to the physical system we are studying,  to 
the creation (annihilation) of either free electrons, or free 
itinerant holons or fermionic atoms in one of two possible hyperfine 
configurations, denoted  symbolically by $\sigma=\uparrow$ and 
$\sigma=\downarrow$. The energy $\varepsilon_{\bf k}$ of those 
fermions is measured with respect to the chemical potential $\mu$.  
Correspondingly, $b_{\bf q}^{\dagger}$ ($b_{\bf q}$) refer to bound 
diatomic molecules of bosonic character (either localized bipolarons, 
or bound hole pairs on plaquette RVB states or weakly bound pairs 
of atoms in a triplet configuration), having an energy $E_{\bf q}$ 
being measured with respect to  $2\mu$. The parameter $2\nu = 
 E_{q=0} - 2\varepsilon_{{\bf k}_F}$ (where ${\bf k}_F$ is 
the Fermi momentum), denotes the difference in energy of the weakly 
bound fermion pairs and the single fermion scattering states. If 
$\nu$ is small, pairing will be introduced among the uncorrelated 
fermions via  resonance scattering, tantamount to a boson-fermion
pair exchange with coupling strength $v$. Tuning the value of $\nu$, 
one can cover the whole regime between Cooper pairs and locally bound 
pairs and their corresponding condensed phases.

Such a BFM (\ref{BF}) has been introduced originally in solid state 
theory many years ago, in an attempt to describe the situation of 
intermediary  electron - lattice coupling \cite{Ranninger-85} and 
has been intensively studied over the last decade, mainly in connection 
with the pseudogap phenomenon in the HTSC. As shown recently 
\cite{PhysRevA-03}, this model does indeed capture the resonant-type 
scattering between fermions due to the Feshbach mechanism and has been 
widely studied in connection with several issues of the atomic gas 
superfluidity \cite{Feshbach-BFM}. 

Our main objective here is to study the two-fermion dynamical correlation 
functions when the {\it detuning} $\nu$ from the resonance is small, thus 
putting ourselves in the center of the cross-over regime between a superfluid 
ground state of BCS characteristics and one corresponding to tightly bound 
fermion pairs of bosonic character. The Green's function describing 
the fermion pairs $G^{pair}({\bf q},\omega)$ is  related to the single 
particle boson propagator via $G^{B}({\bf q},\omega) 
= G^{B}_{0}({\bf q},\omega) + v^{2} G^{B}_{0}({\bf q},
\omega) G^{pair}({\bf q},\omega) G^{B}_{0}({\bf q},\omega)$, 
where $G^{B}_{0}({\bf q},\omega)=\left[ \omega - E_{\bf q} 
-2\nu \right]^{-1}$. This implies that the excitation 
energies of the bound molecules and fermionic diatomic pairs are 
{\em identical}. Only the spectral weights differ as can be seen 
from the relation between their spectral functions, i.e.,  
$A^{pair}({\bf q},\omega)  =  v^{-2} \left( \omega - E_{\bf q} 
- 2\nu \right)^{2} \;  A^{B}({\bf q},\omega)$. It is thus sufficient 
to determine one of these functions in order to derive the excitation 
spectra for both.

\section{The procedure} 
The inter-dependence between the single and two-particle correlations 
requires to treat them on equal footings. For that purpose we  employ 
a continuous renormalization group procedure\cite{Wegner-94} which, 
through a set of infinitesimal canonical transformations, reduces 
the initial Hamiltonian (\ref{BF}) to an essentially diagonalizable form, 
containing the relevant physics which we want to describe, plus additional 
terms which can be treated as small perturbations. Contrary to standard  
renormalization group techniques, where one integrates out the high energy 
states and subsequently derives an effective low energy Hamiltonian, in this 
method both, the high and low energy sectors, are renormalized and kept 
throughout the whole transformation process.

The specific construction of such a procedure for the BFM was given 
previously \cite{Domanski-01}, where also the single particle spectrum 
of the fermionic atoms, pointing to  Bogoliubov-like excitations below 
as well as above $T_{c}$  was studied \cite{PRL-03}. We apply here this 
procedure for the  study of the boson spectral function. In the course of 
diagonalizing the Hamiltonian, the boson operators evolve toward a form 
given by
\begin{eqnarray}
\tilde{b}_{\bf q} = \tilde{\A}_{\bf q} \; b_{\bf q} + 
\frac{1}{\sqrt{N}} \sum_{\bf k} \tilde{\B}_{{\bf q},{\bf k}}
\; c_{{\bf k}\downarrow} c_{{\bf q}-{\bf k}\uparrow},\;\; 
\tilde{b}^{\dagger}_{\bf q}=(\tilde{b}_{\bf q})^{\dagger}.
\label{b_Ansatz}
\end{eqnarray}
The two complex coefficients appearing in (\ref{b_Ansatz}) 
are  calculated in the limit of the convergence of the 
renormalization flow procedure $\lim_{l=\infty} 
\A_{\bf q}(l)=\tilde{\A}_{\bf q}$ and $\lim_{l=\infty} 
\B_{{\bf q},{\bf k}}(l) = \tilde{\B}_{{\bf q},{\bf k}}$, 
where $l$ denotes the continuous flow parameter. We base 
ourselves on the general relations which describe 
the evolution of operators\cite{PRL-03}
\begin{eqnarray}
dO(l)/dl = [ \eta(l),O(l) ] 
\label{flow}
\end{eqnarray}
with $\eta$ being judiciously chosen \cite{Wegner-94} as
\begin{eqnarray}
\eta(l)= \frac{1}{\sqrt{N}} \sum_{{\bf k},{\bf p}}
\alpha_{{\bf k},{\bf p}}(l) \left(  
c_{{\bf p}\uparrow}^{\dagger} c_{{\bf k}
\downarrow}^{\dagger} b_{{\bf p}+{\bf k}}
- \mbox{h.c.} \right) ,
\label{eta}
\end{eqnarray}
and $\alpha_{{\bf k},{\bf p}}(l)=\left[ \varepsilon_{\bf k}(l) 
+ \varepsilon_{\bf p}(l) - E_{{\bf k}+{\bf p}}(l) \right] 
v_{{\bf k},{\bf p}}(l)$ \cite{Domanski-01}. Coefficients 
$A_{\bf q}(l)$ and $B_{{\bf q},{\bf k}}(l)$ satisfy then 
the renormalization equations
\begin{eqnarray}
\frac{d\A_{\bf q}(l)}{dl} & = &  - \frac{1}{N} 
\sum_{\bf k} \alpha_{{\bf k},{\bf q}-{\bf k}}(l)
\; f_{{\bf k},{\bf q}-{\bf k}} \;
\B_{{\bf q},{\bf k}}(l) ,
\label{A_flow}
\\
\frac{d\B_{{\bf q},{\bf k}}(l)}{dl} & = & 
\alpha_{{\bf k},{\bf q}-{\bf k}}(l)  \A_{\bf q}(l) ,
\label{B_flow}
\end{eqnarray}
with the initial conditions $\A_{\bf q}(0)=1$, 
$\B_{{\bf q},{\bf k}}(0)=0$ and $f_{{\bf k},{\bf p}} = 
1 - n_{{\bf k}\downarrow}^{F} - n_{{\bf p}\uparrow}^{F}$.

This procedure leads finally to the following form of the spectral 
function for the bosonic molecules
\begin{eqnarray}
A^{B}({\bf q},\omega) & = & | \tilde{\A}_{\bf q}|^{2} 
\delta \left( \omega - \tilde{E}_{\bf q} \right) +
\label{b_spectral} \\
& & \frac{1}{N} \sum_{\bf k} f_{{\bf k},{\bf q}-{\bf k}}
| \tilde{\B}_{{\bf q},{\bf k}} |^{2} 
\delta \left( \omega - \tilde{\varepsilon}_{\bf k}
- \tilde{\varepsilon}_{{\bf q}-{\bf k}} \right). 
\nonumber
\end{eqnarray}
The first term of (\ref{b_spectral}) describes long-lived 
quasi-particles with the renormalized energy $\tilde{E}_{\bf q}$ 
\cite{Domanski-01} and whose spectral weight is $| \tilde{\A}_{\bf 
q}|^{2}$. The second term describes the incoherent background 
extending over the region determined  by the renormalized fermion 
energies $\tilde{\varepsilon}_{\bf k}$ \cite{Domanski-01}.
From equations (\ref{A_flow},\ref{B_flow}) we derive the following 
sum rule $| \tilde{\A}_{\bf q}|^{2} + \frac{1}{N} \sum_{\bf k} 
| \tilde{\B}_{{\bf q},{\bf k}} |^{2} f_{{\bf k},{\bf q}-{\bf k}} = 1$
which correctly preserves the total spectral weight $\int_{-\infty}
^{\infty} d\omega A^{B}({\bf q},\omega)=\langle [ b_{\bf q}, 
b_{\bf q}^{\dagger} ] \rangle=1$. 

\begin{figure}
\centerline{\epsfxsize=7cm \epsffile{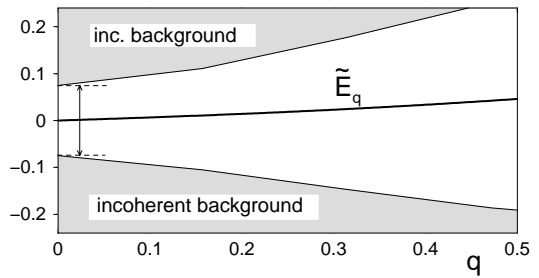}}
\caption{The characteristic sound-wave mode $\tilde{E}_{\bf q}=v|{\bf q}|$
of the long-lived boson and/or fermion pair excitation spectrum in the
superfluid state at $T=0$. The shaded regions show the incoherent background, 
which is energetically separated from the collective excitation branch. 
At ${\bf q}={\bf 0}$ the incoherent background exists for energies larger
than $2 v\chi_{B}$ (twice the value of the single particle fermion gap).
We used the following dispersions $\varepsilon_{\bf k}=-\frac{D}{2} 
\mbox{cos}(a|{\bf k}|)$ and $E_{\bf q}= -\frac{D}{4} \mbox{cos}(a|{\bf q}|)$ 
such that the mass $m_{B}=2m_{F}$ and the potential $v=0.1D$. 
We further set the lattice constant $a=1$ and use the bandwidth $D$ 
as a unit for energies.}
\label{Fig1}
\end{figure}

\section{The pair excitation spectrum below $T_{c}$} 

At a certain critical temperature $T_{c}$ the static pair susceptibility 
$\sum_{{\bf k},{\bf p}} \int_{0}^{\beta} d \tau e^{\tau \omega} \langle 
c_{{\bf k}\uparrow}^{\dagger}(\tau) c_{{\bf q}-{\bf k}\downarrow}^{\dagger}
(\tau) c_{{\bf q}-{\bf p} \downarrow} c_{{\bf p}\uparrow}\rangle_{|\omega 
\rightarrow 0}$ becomes divergent for ${\bf q}={\bf 0}$ and, due to the 
Thouless criterion, the system undergoes a phase transition to a superfluid 
state. For $T<T_{c}$ there appear two order parameters which are proportional 
to each other: $\chi_F \equiv \langle c_{-{\bf k}\downarrow} 
c_{{\bf k}\uparrow}\rangle$ for the fermions and $\chi_{B} \equiv \langle 
b_{{\bf q}={\bf 0}} \rangle$ for the bosons (atom molecules).

Near the Fermi energy, the single particle fermionic excitations 
become  gaped: $\tilde{\varepsilon}_{\bf k}=\mbox{sgn} \left\{ 
\varepsilon_{\bf k} \right\}\;\sqrt{(\varepsilon_{\bf k})^{2}
+(v\chi_{B})^{2}}$. In consequence, no fermionic states, neither 
coherent nor incoherent, exist within the energy window $|\omega| 
\leq v\chi_{B}$ \cite{PRL-03}. This simultaneously affects the 
incoherent part of bosonic spectrum, as can be seen from (\ref{b_spectral}). 
For the long wavelength limit ${\bf q} \rightarrow {\bf 0}$ the incoherent 
background is pushed up to energies $|\omega| > 2 v \chi_{B}$ and thus 
permits long-lived excitations, which correspond to collective modes, known 
as {\em first sound} for interacting bosonic systems in the superfluid 
state (see Fig.1). The temperature dependence of these modes has previously 
been studied for this BFM \cite{Tomek-96} in the superfluid phase within 
a framework of the dielectric formalism with use of the Ward identities, 
currently employed in the theory of interacting bose gases. A behavior 
similar to that of the strong coupling limit of interacting bose gases 
\cite{Szepfalusy-74} was found, showing a sound velocity being little 
dependent on temperature as one traverses the superfluid transition, 
but whose spectral weight in the boson single particle spectral 
function disappears upon approaching $T_c$.

Such sound wave-like modes are not realized in charged superconducting 
systems because of the long range Coulomb interaction which pushes them 
up to the generally huge plasma frequency \cite{Anderson-63}. For electrically 
neutral atoms, such as the trapped atomic gases, this is no longer the case 
and hence one can realistically expect collective sound wave-like modes, 
although appropriately modified due to the inhomogeneous structure of 
the gas density \cite{Collmode}.

\section{The pair excitation spectrum above $T_{c}$} 

\begin{figure}
\centerline{\epsfxsize=7.5cm \epsffile{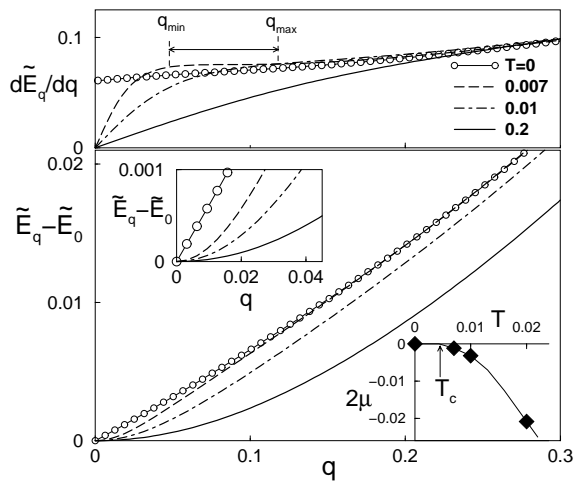}}
\caption{Comparison of the dispersion $\tilde{E}_{\bf q}$ of
the coherent part of the boson spectral function at temperatures
corresponding to the superfluid ($T=0$), pseudogap ($0.007$, $0.01$)
and the normal phase above $T^{*}$ ($0.02$). Upper panel shows
the derivative $d\tilde{E}_{\bf q}/dq$ of these curves. The insets
contain correspondingly: the low momentum $q$ limit of the dispersion
$\tilde{E}_{\bf q}$ and the temperature dependence of chemical
potential (the marked points correspond to four temperatures
$T=0.02$, $0.01$, $0.007$ and $0$ chosen in this work).}
\label{Fig2}
\end{figure}

Decreasing the temperature in the normal state below a certain 
$T^* (>T_c)$ one expects precursor pairing effects which show up in the  
single particle fermionic excitations spectrum in form of a pseudogap 
which opens up near the chemical potential 
\cite{Domanski-01,PRL-03,Ranninger-95}.  
Above $T^*$ the low energy part of the pair excitations has  the usual 
parabolic dispersion. However, upon decreasing the temperature and 
approaching $T_c$, phase coherence gradually sets in on a finite length 
and time scale, which becomes visible in form of a linear in $q$ 
dispersion of the single boson (respectively Fermion pair) excitation
for small $q$ vectors, in an interval $[q_{min}(T),q_{max}(T)]$ 
(see Fig. 2). There, the derivative of the effective Bose single particle  
energy spectrum $d\tilde{E}_q/dq$ shows a flat portion, which, when 
extrapolated to $q=0$, practically coincides with the corresponding 
quantity  in the superfluid phase at $T=0$. We observe that, as the 
temperature is decreased, $q_{min}(T)$ decreases toward zero, but 
always leaving a small interval in q space $[0,q_{min}(T)]$ where 
one clearly observes a free particle like spectrum with an effective 
mass which decreases as T decreases. This is in accordance with 
an earlier study on this subject using selfconsistent perturbation 
theory\cite{Ranninger-95}. For $T \geq T^*$ the coherent boson mode 
overlaps with an incoherent background in the single particle boson 
spectral function (see Fig.3, bottom panel). Yet, upon decreasing 
the temperature to below $T^*$, we observe that this incoherent 
background moves away from the position of the coherent contribution 
(upper panel of Fig.3) which ensures that a linear in $q$ branch 
of the boson spectrum is well defined in the corresponding interval 
of $q$ vectors. This strongly suggests that remnants of the first 
sound still can exist as part of the single particle boson spectrum 
above the superfluid phase transition for a limited region of wave 
vectors due to a persistence  of superfluid phase correlations 
above $T_c$ on a finite length and time scale.

\begin{figure}
\centerline{\epsfxsize=7cm \epsffile{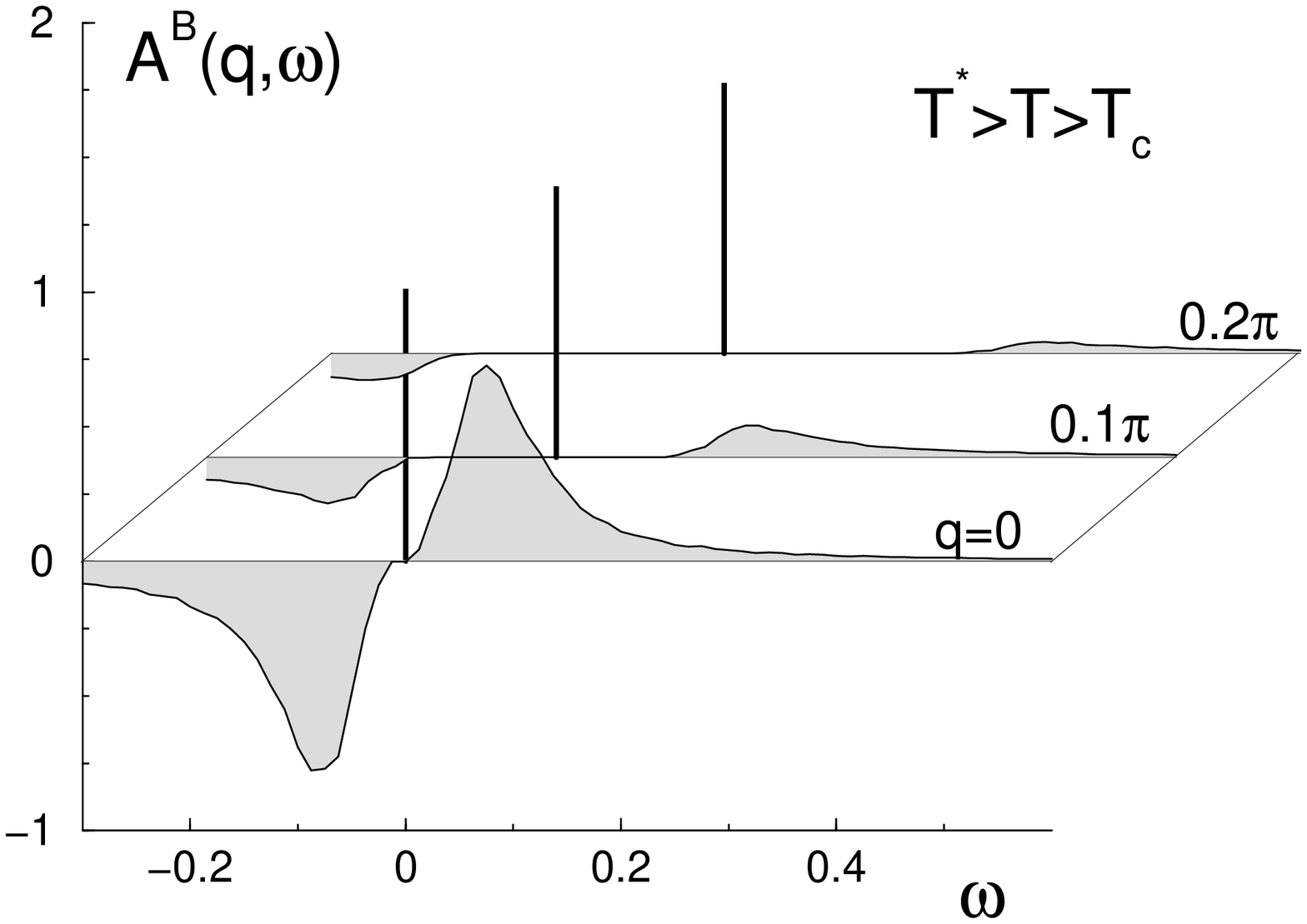}}
\centerline{\epsfxsize=7cm \epsffile{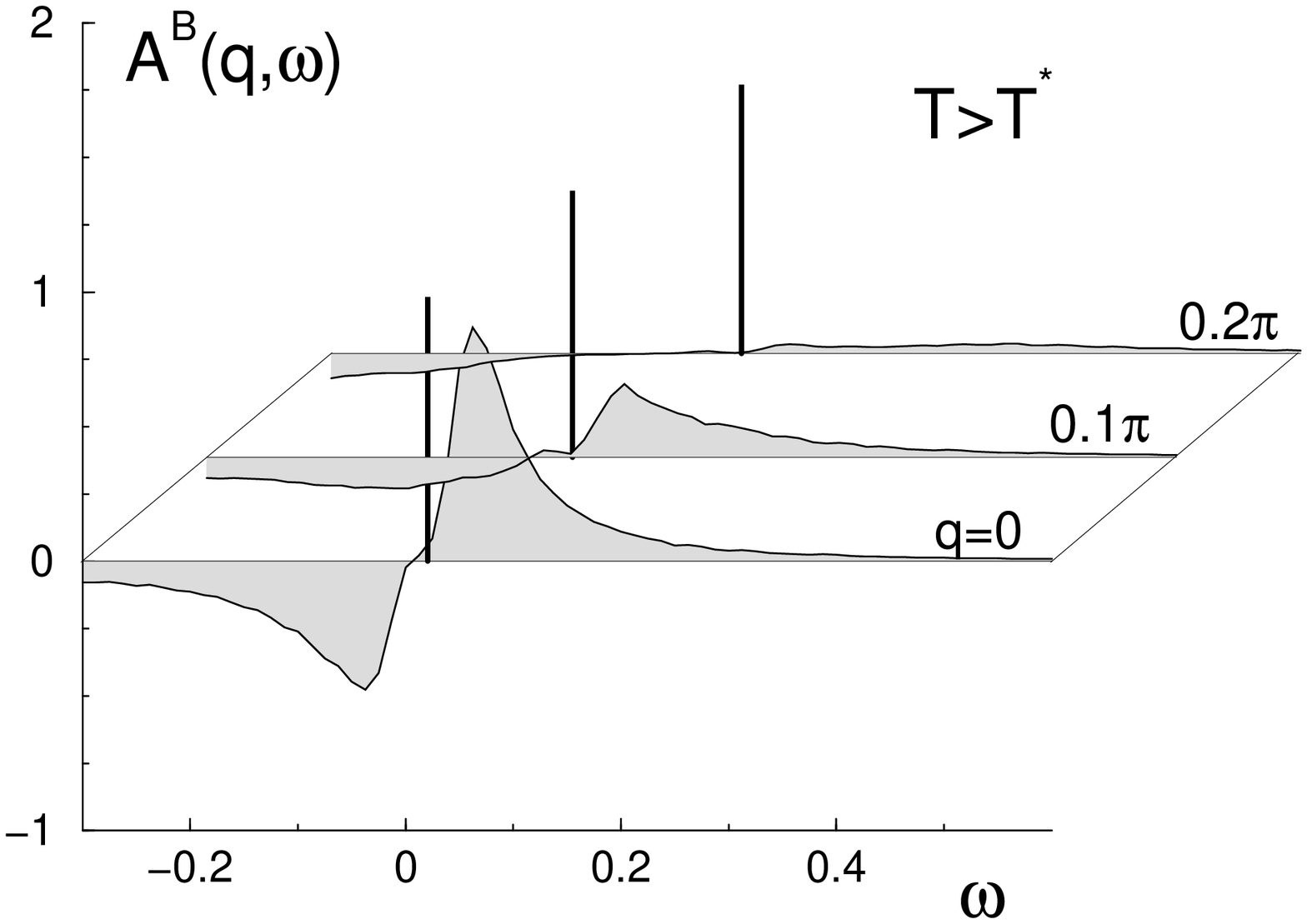}}
\caption{The boson spectral function $A^{B}({\bf q},\omega)$ for the low 
energy pair excitations. The upper panel corresponds to $T=0.007$ being 
close to $T_{c}$ in the pseudogap region $T^{*}>T>T_{c}$. The bottom 
panel refers to the normal state $T=0.02$ (above $T^{*}$). In the pseudogap 
phase a propagating coherent contribution given by the $\delta$-function 
peak and an incoherent background, given by the shaded regions, get separated 
above some relatively small critical momentum $q_{min}$. This is no longer 
the case for $T>T^{*}$.}
\label{Fig3}
\end{figure}  

\section{Conclusions} 

We studied the qualitative changes of the excitation spectrum for 
the resonant fermion pairs which occur upon varying the temperature. 
We found that quantum fluctuations play a crucial role when detuning 
$\nu$ from the Feshbach resonance is small. Fluctuations manifest 
themselves in the pseudogap regime $T^{*}>T>T_{c}$.

Far above $T_{c}$ the off-diagonal long range order is not 
established. The pair excitation spectrum for small $q$ vectors 
is then characterized  by a parabolic branch (see Fig. 2) and  
overlaps with the incoherent background (see the bottom panel of Fig. 3) 
such as to effectively destroy any bosonic quasi-particle features.
 
This situation changes dramatically when the temperature drops below 
$T^{*}$ where resonant pairing sets in. Phase correlations start to 
build up on a finite spatial and temporal scale as the temperature 
decreases and approaches $T_c$. The single particle 
fermion spectrum reveals then a partial suppression of states 
(pseudogap) around the Fermi energy \cite{Domanski-01,PRL-03,Ranninger-95}, 
which is accompanied by qualitative changes in the pair excitation spectrum. 
Quantum fluctuations lead to emergence of the collective sound-wave 
mode which above $T_{c}$ exists in a finite  momentum interval 
$[q_{min}(T),q_{max}(T)]$. Upon decreasing the temperature 
the long-lived branch of the pair spectrum gradually splits off 
from the incoherent background (upper panel in Fig. 3) and spreads 
over a wider and wider momentum region, with $q_{min}(T)$ steadily 
decreasing as we approach $T_c$. We note however that invariably 
the linear in $q$ dispersion changes into a damped free particle 
like behavior in the close vicinity of $q=0$. 

The sound-wave mode has been so far measured above $T_c$ 
in the liquid helium by ultrasonic techniques \cite{ultrasonic} 
as well as by neutron scattering measurements \cite{Glyde-04}. 
In the case of the trapped fermionic atoms the corresponding 
mode is expected to be compressional density waves and, similar 
to the present study, one should expect remnants of those modes 
in the normal state. In principle, such modes can be experimentally 
checked by the Bragg spectroscopy \cite{Steinhauser-02}. Indirect 
methods for detecting the collective modes which rely on measuring 
the magnetic susceptibility and density-density correlation functions 
have been discussed (although only for $T<T_{c}$) in Ref.\ \cite{Ohashi}. 
In some future work we shall discuss how collective modes can possibly 
be observed in measurements of the magnetic susceptibility in  
the pseudogap regime above $T_{c}$.

{\em Acknowledgment.} --
We would like to thank Henry R.\ Glyde for discussions on the analogous 
problems in the liquid helium and Dr.\ Tomasz Kostyrko for valuable 
comments on the general physics  behind this work. T.D.\ kindly
acknowledges hospitality of the Laue Langevin Institute in 
Grenoble where the main  part of this work has been completed. 
T.D.\ was partly supported by the Polish Committee of Scientific 
Research under the grant No.\ 2P03B06225.

\end{document}